\documentclass[journal,12pt,onecolumn]{IEEEtran}

\usepackage{times,bm}
\usepackage{bbold}
\usepackage{bbm}
\usepackage{url}
\usepackage{indentfirst}
\usepackage{cite}
\usepackage{subfigure}
\usepackage{amsmath}
\usepackage{amssymb}
\usepackage{multicol}
\usepackage{multirow}
\usepackage{amsfonts}
\usepackage{ textcomp }
\usepackage{upgreek}
\usepackage{times}
\usepackage[dvips]{graphicx}
\usepackage[usenames]{color}
\usepackage{amsthm}

\newcommand{\ben}{\begin{enumerate}}
\newcommand{\een}{\end{enumerate}}
\newcommand{\be}{\begin{equation}}
\newcommand{\ee}{\end{equation}}
\newcommand{\bal}{\begin{align}}
\newcommand{\eal}{\end{align}}
\newcommand{\bea}{\begin{eqnarray}}
\newcommand{\eea}{\end{eqnarray}}
\newcommand{\bc}{\begin{cases}}
\newcommand{\ec}{\end{cases}}
\newcommand{\bi}{\begin{itemize}}
\newcommand{\ei}{\end{itemize}}

\newtheorem{teo}{\textbf{Theorem}}
\theoremstyle{definition}

\newtheorem{lem}{Lemma}

\begin{document}

\title{Mobile Relays for Smart Cities: Mathematical Proofs}

\author{{\bf Tugcan Aktas}, {\bf Giorgio Quer}, {\bf Tara Javidi}, {\bf Ramesh R. Rao}\\
{
University of California San Diego -- La Jolla, CA 92093, USA.}}

\maketitle

\section{Proofs for S$_M$ under non-stopping fmBS scenario}
\begin{figure}[htbp]
\centering
\includegraphics[width=1\linewidth]{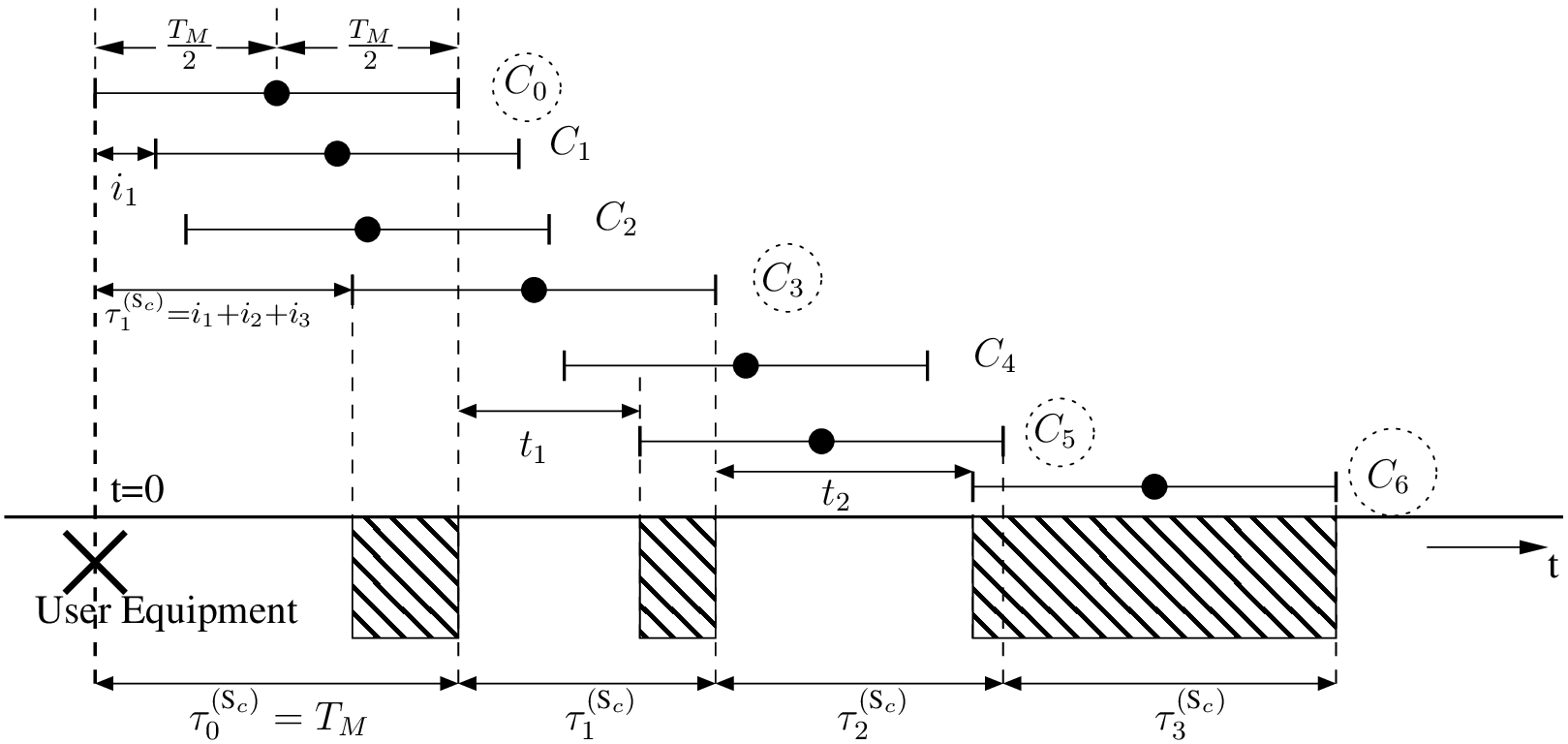}
\caption{Time of arrivals of fmAPs and S$_\text{c}$ operation for a given T2 connection round.}
\label{fig:maxsel}
\end{figure} 
We first present the following two lemmas: 

\begin{lem}{The expectation of service time $\tau^{(\text{S}_c)}_1$ is}
\label{lem:S1}
\begin{align}
\label{eqn:s1_exp2}
\text{E} \left[ \tau^{(\text{S}_c)}_1 \vert M^{(\text{S}_c)}\geq 1 \right] = \frac{\lambda T_{M} - \left( 1-e^{ -\lambda T_{M}}\right)}{\lambda \left( 1-e^{ -\lambda T_{M}}\right)}.
\end{align}
\end{lem}

\begin{lem}{The expectation of $t_1$ is}
\label{lem:Ri}
\begin{align}
\text{E} & \left[ t_1 \vert M^{(\text{S}_c)}\geq 1\right] = \frac{\lambda}{e^{ \lambda T_{M}}-1}\int_0^{T_{M}}\frac{s_1 e^{ \lambda s_1}}{1-e^{\lambda s_1}}\textrm{d} s_1 - \frac{1}{\lambda}\nonumber \\
\label{eqn:r_i_exp2}
&= \frac{1}{\lambda \left( e^{\lambda T_{M}} - 1\right)}\left(\textrm{Li}_2\left( e^{\lambda T_{M}}\right) - \frac{\pi^2}{6} + 1 \right. \\
&+\left.\vphantom{\frac{\pi^2}{6}} e^{\lambda T_{M}}\left(\lambda T_{M} - 1\right) + \lambda T_{M} \log \left( 1-e^{\lambda T_{M}}\right) \right) - \frac{1}{\lambda},\nonumber
\end{align}
\end{lem}
where $\textrm{Li}_2\left(x\right)\triangleq \sum_{k=1}^{\infty}\frac{x^k}{k^2}$ is the polylogarithm function of order $2$. The proofs of the these two lemmas are given consecutively.

\subsection{Proof of Lemma~\ref{lem:S1}}
\label{sec:appS1}
Let us start with the conditional pdf $f_{\tau^{(\text{S}_c)}_1\vert M^{(\text{S}_c)}\geq 1}\left(s\right)$. For $M^{(\text{S}_c)}=0$, we have $\tau^{(\text{S}_c)}_1 = 0$, which is a trivial result. In the case $M^{(\text{S}_c)}\geq 1$, we have $K\geq 1$ fmBS arrivals in the interval $(0,T_{M}]$ by taking the arrival time of the initial vehicle $C_0$ as our time origin. Furthermore, according to the uniform property of the Poisson arrivals, the exact arrival times of these fmBSs, namely $A_j,j\in \left\lbrace 1, \hdots, K \right\rbrace$, are independent and distributed uniformly in $(0,T_{M}]$. Therefore, we have $\tau^{(\text{S}_c)}_1 = \max_{j\in \left\lbrace 1, \hdots, K \right\rbrace} \left\lbrace A_j \right\rbrace$. It is known that the pdf of the maximum of $k$ independent uniform random variables defined in $(0,T_{M}]$ satisfies
\begin{align}
\label{eqn:s1_condK}
f_{\tau^{(\text{S}_c)}_1\big\vert M^{(\text{S}_c)}\geq 1, K=k}\left(s\right) = \frac{k s^{k-1}}{T_{M}^k}, \;\;s\in \left(0,T_{M}\right].
\end{align}
Since $K$ is a Poisson random variable with parameter $\lambda T_{M}$ and we are given that $K\geq 1$, the expectation sum over (\ref{eqn:s1_condK}) yields
\begin{align}
f_{\tau^{(\text{S}_c)}_1\vert M^{(\text{S}_c)}\geq 1}\left(s\right) &= \sum_{k=1}^{\infty}\frac{k s^{k-1}}{T_{M}^k} \frac{\left( \lambda T_{M}\right) ^{k} }{k!}\frac{\exp \left( -\lambda T_{M}\right) }{1-\exp\left( -\lambda T_{M}\right)}\nonumber\\
\label{eqn:s1_exp}
&=\frac{\lambda \exp\left( \lambda s\right)}{\exp\left( \lambda T_{M}\right) -1 }, \;\;s\in \left(0,T_{M}\right],
\end{align}
where we make use of the definition for the probability generating function (PGF) of $K$. By using (\ref{eqn:s1_exp}), one can evaluate the desired expected value for $\tau^{(\text{S}_c)}_1$ by integrating (\ref{eqn:s1_exp}) over $s\in \left(0,T_{M}\right]$.

\subsection{Proof of Lemma~\ref{lem:Ri}}
\label{sec:appRi}
Similar to proof of Lemma~\ref{lem:S1}, for $t_j$ to be different than zero, we require $K\geq 1$ arrivals in the $(0,\tau^{(\text{S}_c)}_j=s_j]$ interval. Then, we have
\begin{align}
\label{eqn:ri_condK}
f_{t_j\vert M^{(\text{S}_c)}\geq j+1, K=k, K\geq 1, \tau^{(\text{S}_c)}_j = s_j}\left(r\right) = \frac{k r^{k-1}}{s_j^k}, \;\;r\in \left(0,s_j\right].
\end{align}
Taking the expectation over the pmf of $K$ with the condition $K\geq 1$, we obtain
\begin{align}
\label{eqn:ri_cond_si}
f_{t_j\vert M^{(\text{S}_c)}\geq j+ 1, \tau^{(\text{S}_c)}_j = s_j}\left(r\right)= \frac{\lambda \exp\left( \lambda r\right)}{\exp\left( \lambda s_j\right) -1 }, \;\;r\in \left(0,s_j\right].
\end{align}
The expected value of $t_j$, conditioned on $\tau^{(\text{S}_c)}_j=s_j$, follows from an expectation integral of (\ref{eqn:ri_cond_si}) over $r\in \left(0,s_j\right]$.

\begin{align}
\label{eqn:r_i_exp}
\text{E} \left[ t_j \vert M^{(\text{S}_c)}\geq j+1, \tau^{(\text{S}_c)}_j = s_j \right] 
&= \int_0^{s_j} r \frac{\lambda \exp\left( \lambda r\right)}{\exp\left( \lambda s_j\right) -1 } \textrm{d}r\nonumber\\
&= \frac{\lambda s_j - \left( 1-\exp\left( -\lambda s_j\right)\right)}{\lambda \left( 1-\exp\left( -\lambda s_j\right)\right)}.
\end{align}

In order to remove the condition on $\tau^{(\text{S}_c)}_1$ for the $j=1$ case and obtain $\text{E} \left[ t_1 \vert M^{(\text{S}_c)}\geq 1\right]$, we first multiply of the expressions in (\ref{eqn:r_i_exp}) and (\ref{eqn:s1_exp}) to further integrate the resulting expression over $s_1\in \left(0,T_{M}\right]$, which finalizes the proof.  

\begin{teo}{The expected effective ratio of time in T2 with strategy S$_\text{c}$ can be approximated by}
\label{thm:exp_time_SM}
%\begin{align}
%\label{eqn:max_sel_handoff}
% R^{(\text{S}_c)}_2 \simeq \left(1-P_V\right) -\frac{2\; \left(1-P_V\right) T_H}{ \left( \text{E} \left[ \tau^{(\text{S}_c)}_1 \vert M^{(\text{S}_c)}\geq 1 \right] + \text{E} \left[ t_1 \vert M^{(\text{S}_c)}\geq 1 \right]\right) + \frac{2}{\lambda} },
%\end{align}
\begin{align}
\label{eqn:max_sel_handoff}
 R^{(\text{S}_c)}_2 \simeq \left(1-P_V\right) -\frac{2\; \left(1-P_V\right) T_H}{  \text{E} \left[ \tau^{(\text{S}_c)}_1 + t_1 \big\vert M^{(\text{S}_c)}\geq 1\right] + \frac{2}{\lambda} },
\end{align}
where $\text{E} \left[ \tau^{(\text{S}_c)}_1 \vert M^{(\text{S}_c)}\geq 1 \right]$ follows from (\ref{eqn:s1_exp2}) and $\text{E} \left[ t_1 \vert M^{(\text{S}_c)}\geq 1 \right]$ from (\ref{eqn:r_i_exp2}).
\end{teo}

\subsection{Proof of Theorem~\ref{thm:exp_time_SM}}
\label{sec:corrN}

For a given value $M^{(\text{S}_c)}=n$, the time intervals in which new fmBS arrivals occur are disjoint time intervals of length $\tau^{(\text{S}_c)}_1, t_1, t_2, \hdots, t_{n-1}$, as exemplified in Fig.~\ref{fig:maxsel}. With an abuse of notation, we drop the superscript $(\text{S}_c)$ in $\tau^{(\text{S}_c)}_1$ since for no strategy other than S$_M$, we deal with $\tau^{(\text{S})}_j$ in the scope of this proof. Therefore, the number of unserved fmBSs, $U$, in a T2 round satisfies
\begin{align}
\text{E} \left[ U \vert M^{(\text{S}_c)}=n \right] = \text{E} \left[ U^{\tau_1} + \sum_{j = 1}^{n-1} U^{t_j} \right], \nonumber
\end{align}
where $U^{\tau_1}$ and $U^{t_j}$ are the number of unserved fmBSs in the time intervals $\tau_1$ and $t_j$, respectively. Based on the iterated expectations over these random service times, we obtain
\begin{align}
&\text{E} \left[ U \vert M^{(\text{S}_c)}=n \right] {=} \text{E} \left[ \text{E}_{
U\vert \tau_1,t_j,M^{(\text{S}_c)}=n} \left[ U^{\tau_1} {+} \sum_{j = 1}^{n-1}  U^{t_j}\right] \right]\nonumber\\
\label{eqn:unserved_exp2}
&=\lambda\left( \text{E} \left[ \tau_1 \vert M^{(\text{S}_c)}=n \right] + \sum_{j = 1}^{n-1} \text{E} \left[ t_j \vert M^{(\text{S}_c)}=n \right]\right),
\end{align}
where we use the fact that $\text{E} \left[ U^{\tau_1}\vert \tau_1 = s_1\right] = \lambda s_1$, and a similar argument is valid for $t_j$ as well. 

We should evaluate $\text{E} \left[ \tau_1 \vert M^{(\text{S}_c)}=n \right]$ and $\text{E} \left[ t_j \vert M^{(\text{S}_c)}=n\right]$ in order to remove the conditions on these random variables in (\ref{eqn:unserved_exp2}). However, it is possible to approximate the term $\text{E} \left[ U \vert M^{(\text{S}_c)}=n \right]$ by evaluating $\text{E} \left[ \tau_1 \vert M^{(\text{S}_c)}\geq 1 \right]$, and for only a few values of $\text{E} \left[ t_j \vert M^{(\text{S}_c)}\geq j+1 \right]$. %In the ideal case with steady state assumption on the number of unserved fmBSs in S$_\text{c}$ handoff, one would expect that $\text{E} \left\lbrace R_i \vert N\geq i+ 1 \right\rbrace$ would converge to a single value for $i\rightarrow \infty$. However, for the initial serving time, $\text{E} \left\lbrace S_1 \vert N\geq 1 \right\rbrace$ might strictly deviate from this expectation since it follows a deterministic service time $T_{M}$ of the initiator fmBS and observes a transient state for the investigated queue. 
Therefore, we use 

\begin{align}
\label{eqn:unserved_exp_apprx}
\text{E} \left[ U \vert M^{(\text{S}_c)}{=}n \right] {\simeq}  n \lambda\frac{\text{E} \left[ \tau_1 \vert M^{(\text{S}_c)}{\geq} 1 \right] {+} \text{E} \left[ t_1 \vert M^{(\text{S}_c)}{\geq} 1 \right]}{2},
\end{align}
where the term $\left(\text{E} \left[ \tau_1 \vert M^{(\text{S}_c)}\geq 1 \right] + \text{E} \left[ t_1 \vert M^{(\text{S}_c)}\geq 1 \right]\right)/2$ is an approximation for the average number of unserved fmBSs between two consecutive horizontal handoffs. %Moreover, one can remove the condition on $S_1$ in Lemma~\ref{lem:Ri} by using (\ref{eqn:s1_exp}).

On the other hand, the number of handoffs for S$_\text{m}$ and S$_\text{c}$ in the whole T2 round satisfy
\begin{align}
\label{eqn:all_exp_apprx}
&\text{E} \left[ M^{(\text{S}_m)} \right] = \text{E}_{M^{(\text{S}_c)}} \left[ \text{E} \left[ U + M^{(\text{S}_c)}\vert M^{(\text{S}_c)} \right] \right] \nonumber\\
 & {\simeq} \text{E} \left[ M^{(\text{S}_c)}\right] \left( \lambda\frac{\text{E} \left[ \tau_1 \vert M^{(\text{S}_c)}{\geq} 1 \right] {+} \text{E} \left[ t_1 \vert M^{(\text{S}_c)}{\geq} 1 \right]}{2}{+}1\right),
\end{align}
where we use (\ref{eqn:unserved_exp_apprx}). Solving it for $\text{E} \left[ M^{(\text{S}_c)}\right]$ in (\ref{eqn:all_exp_apprx}) we obtain
\begin{align}
\label{eqn:max_sel_M}
 \text{E} \left[ M^{(\text{S}_c)}\right] \simeq \frac{2\; \text{E} \left[ M^{(\text{S}_m)}\right]}{\lambda \left( \text{E} \left[ \tau^{(\text{S}_c)}_1 + t_1 \vert M^{(\text{S}_c)}\geq 1 \right]\right) + 2 },
\end{align}
where the denominator follows from the results of Lemmas~\ref{lem:S1} and~\ref{lem:Ri}. For S$_\text{c}$, the expected effective ratio of time in T2 is
\begin{align}
\label{eqn:max_sel_ratio}
R^{(\text{S}_c)}_2 = \frac{\text{E} \left[ T_2 \right] - \text{E} \left[ M^{(\text{S}_c)} \right] T_H}{\text{E} \left[ T_1 \right]+\text{E} \left[ T_2 \right]},
\end{align}
where $\text{E} \left[ T_1 \right]=\lambda^{-1}$ and $\text{E} \left[ T_2 \right]=\frac{ 1 - P_V}{\lambda P_V}$ as for S$_m$~\cite{GC_16}. The proof is completed by plugging all these known expressions and (\ref{eqn:max_sel_M}) into (\ref{eqn:max_sel_ratio}).

\section{Proofs for S$_m$ under stopping fmBS scenario}
\begin{lem}
\label{lem:M_stop}
\begin{align}
\label{eqn:M_stop}
\textrm{E}\left[ M^{(\text{S}_m)}\right] =\sum_{m=0}^{\infty}\prod_{j=1}^{m+1}\left(1-\hat{P}_{V}^{(j)}\right),
\end{align}
where $\hat{P}_{V}^{(j)}$ is the probability of a vertical handoff at the end of a service time of the $(j-1)^{\text{st}}$ fmBS and is expressed as
\begin{align}
&\hat{P}_{V}^{(j)} = e^{-\lambda T_{M}}\left[1-P_S\left(1-e^{-P_S\lambda T_S}\right) \frac{1-\Delta^j}{1-\Delta}\right]
\end{align}
and $\Delta \triangleq \left(P_S'-P_S\right)$.
\end{lem}

\subsection{Proof of Lemma~\ref{lem:M_stop}}
\label{sec:app_M_stop}
The conditional probability that the $j^\text{th}$ fmBS is a stopping one, given that the $(j-1)^\text{st}$ fmBS is a stopping one, is
\begin{align}
P_S' \triangleq \left( 1 - e^{-P_S\lambda T_S}\right)+ e^{-P_S\lambda T_S}\left(1-e^{-\lambda T_{M}}\right)P_S,
\end{align}
whereas the probability of observing a stopping one following a non-stopping one is simply $P_S$. Let us define the probability $P_S'^{(j)}\triangleq P\left\lbrace \textrm{$j^\text{th}$ fmBS has stopped} \right\rbrace$. It can be found that $P_S'^{(j)}=P_S\sum_{k=0}^{j} \left( P_S' -P_S\right)^k$ for $j=1,2,\hdots$. Therefore, following several steps of derivation, one reaches the result that 
\begin{align}
&\hat{P}_{V}^{(j)} = P_S'^{(j-1)} \text{P}\left\lbrace N_{ns}\left( 0, T_{M}\right] + N_{s}\left( 0, T_{M}+T_S\right] = 0\right\rbrace\nonumber\\
&+ ( 1-P_S'^{(j-1)}) \text{P}\left\lbrace N_{ns}\left( 0, T_{M}\right] + N_{s}\left( 0, T_{M}\right] = 0\right\rbrace\nonumber\\
&= P_S'^{(j-1)}\text{P}\left\lbrace N_{ns}\left( 0, T_{M}\right] = 0\right\rbrace \text{P}\left\lbrace  N_{s}\left( 0, T_{M}+T_S\right] = 0\right\rbrace\nonumber\\
%\label{eqn:indep_process}
&+ ( 1-P_S'^{(j-1)}) \text{P}\left\lbrace N_{ns}\left( 0, T_{M}\right] = 0\right\rbrace \text{P}\left\lbrace N_{s}\left( 0, T_{M}\right] = 0\right\rbrace\nonumber\\
\label{eqn:P_i_2_1}
&=e^{-\lambda T_{M}}\left[1-P_S\left(1-e^{-P_S\lambda T_S}\right) \frac{1-\Delta^j}{1-\Delta}\right],
\end{align}
where $N_{s}\left(.,.\right]$ and $N_{ns}\left(.,.\right]$ are the independent counting processes corresponding to the Poisson processes of stopping and the non-stopping fmBS arrivals as defined in~\cite{GC_16}, and $\Delta \triangleq \left(P_S'-P_S\right)$. Therefore, one can evaluate the expected value for the T2 handoffs for the stopping fmBS case utilizing S$_m$ as
\begin{align}
\label{eqn:expected_M_stopping}
\textrm{E}\left[ M^{(\text{S}_m)}\right] &=\sum_{m=0}^{\infty}m\prod_{j=1}^{m}\hat{P}_{V}^{(m+1)}\left(1-\hat{P}_{V}^{(j)}\right),\nonumber\\
&=\sum_{m=0}^{\infty}\prod_{j=1}^{m+1}\left(1-\hat{P}_{V}^{(j)}\right),
\end{align}
where we used the fact that for any non-negative random variable $M$, $\textrm{E}\left[ M\right]=\sum_{m=0}^{\infty} \textrm{P}\left( M > m\right)$. Since the expression in (\ref{eqn:expected_M_stopping}) is an infinite summation, one can either directly utilize it to approximate $\textrm{E}\left[ M^{(\text{S}_m)}\right]$ by truncating the summation at a finite $m$ value, or can model random variable $M^{(\text{S}_m)}$ using a new random variable $M'$ which has the following success probabilities 
\begin{align}
P_{M'}^{(1)} &= \hat{P}_{V}^{(1)} = e^{-\lambda T_{M}}\left[1-P_S\left(1-e^{-P_S\lambda T_S}\right) \right]\nonumber\\
P_{M'}^{(m)} &= \hat{P}_{V}^{(2)}\nonumber\\
&= e^{-\lambda T_{M}}\left[1-P_S\left(1-e^{-P_S\lambda T_S}\right)\left(1+P_S'-P_S\right) \right],\nonumber
\end{align}
where $m\geq 2$. Using this model we obtain
\begin{align}
\textrm{E}\left[ M^{(\text{S}_m)}\right] &\simeq \textrm{E}\left[ M' \right] = \frac{1-P_{M'}^{(1)}}{P_{M'}^{(2)}}\nonumber\\
\label{eqn:M_approx}
&=\frac{1-e^{-\lambda T_{M}}\left[1-P_S\left(1-e^{-P_S\lambda T_S}\right)\right]}{e^{-\lambda T_{M}}\left[1-P_S\left(1-e^{-P_S\lambda T_S}\right)\left(1+P_S'-P_S\right) \right]}.
\end{align}

\begin{teo}{The expected effective T2 ratio for S$_\text{m}$ with stopping fmAPs can be approximated by}
\label{thm:exp_T2_stop_approx}
\begin{align}
\label{eqn:ratio_stopping}
\hat{R}^{(S_m)}_2 \simeq \frac{\tilde{A}_2-T_H \text{E}\left[ M^{(S_m)}\right]}{\text{E} \left[\hat{T}_1\right]+\tilde{A}_2},
\end{align} 
\end{teo}
\noindent
where $\textrm{E}\left[\hat{T}_1\right]=1/\lambda$, $\text{E}\left[ M^{(S_m)}\right]$ is given in (\ref{eqn:M_approx}), and we approximate $\text{E} \left[ \hat{T}_2 \right] \simeq \tilde{A}_2$, which is defined as
\begin{align}
\tilde{A}_2 =  T_M {+} P_S T_S {+} \text{E} \left[ M^{(S_m)}\right] \frac{\textrm{E}\left[ \tau^{(S_m)}_1 \right] {+} \textrm{E}\left[ \tau^{(S_m)}_2 \right]}{2}.
\end{align}

\subsection{Proof of Thm.~\ref{thm:exp_T2_stop_approx}}
\label{sec:app_T2_stop}

The expected service time $\textrm{E}\left[ \tau^{(\text{S}_m)}_j \right]$ can be evaluated by using $4$ possible combinations of the stopping property for the $(j-1)^{\text{st}}$ and the $j$th fmAPs and it simplifies to
\begin{align}
\label{eqn:exp_I_i}
&\textrm{E}\left[ \tau^{(\text{S}_m)}_j \right] = \left(1-P_S'^{(j-1)}\right) \frac{1-e^{-x}\left(1+x\right)}{\lambda\left(1-e^{-x}\right)}{\bigg\vert}_{x=P_S\lambda T_{M}}\nonumber \\
&+ \left(1-P_S'^{(j-1)}\frac{P_S'-P_S}{1-P_S}\right) \frac{1-e^{-x}\left(1+x\right)}{\lambda\left(1-e^{-x}\right)}{\bigg\vert}_{x=\left(1-P_S\right)\lambda T_{M}}\nonumber \\
&+P_S'^{(j-1)}\frac{P_S'}{P_S}\frac{1-e^{-x}\left(1+x\right)}{\lambda\left(1-e^{-x}\right)}{\bigg\vert}_{x=P_S\lambda\left( T_{M} + T_S\right)},
\end{align}
where $P_S'\triangleq \left( 1 - e^{-P_S\lambda T_S}\right)+ e^{-P_S\lambda T_S}\left(1-e^{-\lambda T_{M}}\right)P_S$ is the probability that an fmAP is a stopping one given that the previous one has stopped and $P_S'^{(j)} \triangleq P_S\sum_{k=0}^{j} \left( P_S' -P_S\right)^k$ is the probability that the $(j-1)^{\text{st}}$ fmAP is a stopping one.
 
Similar to the approximation in the proof of Thm.~\ref{thm:exp_time_SM}, we can approximate the expected time in T2 by using only a few of the $\textrm{E}\left[ \tau^{(\text{S}_m)}_j \right]$ terms. As an example, only by utilizing $\tau^{(\text{S}_m)}_1$ and $\tau^{(\text{S}_m)}_2$ we reach
\begin{align}
\label{eqn:min_sum_stop}
\text{E} \left[ \hat{T}_2 \right] {\simeq} T_{M} + P_S T_S +\text{E} \left[ M^{(\text{S}_m)}\right] \frac{\text{E}\left[ \tau^{(\text{S}_m)}_1 \right] {+} \textrm{E}\left[ \tau^{(\text{S}_m)}_2 \right]}{2},
\end{align}
where $\textrm{E}\left[ \tau^{(\text{S}_m)}_1 \right]$ follows from (\ref{eqn:exp_I_i}) with $j=1$. The average time spent in T1 is not affected by the stopping fmAPs since a handoff from T1 to T2 occurs only when the next arrival is observed. Hence $\text{E} \left[ \hat{T}_1\right] = 1/\lambda$. The proof is completed when we replace $\text{E} \left[ T_2 \right]$ in both the numerator and the denominator of (\ref{eqn:ratio_stopping}) with the approximation for $\text{E} \left[ \hat{T}_2 \right]$ in (\ref{eqn:min_sum_stop}) and use the result on $\text{E} \left[ M^{(\text{S}_m)}\right]$ from Lemma~\ref{lem:M_stop}.

\bibliographystyle{IEEEtran}
\bibliography{biblio}

\end{document}